\documentstyle[12pt]{article}
\input{psfig}

\def\mincir{\raise -2.truept\hbox{\rlap{\hbox{$\sim$}}\raise5.truept
\hbox{$<$}\ }}
\def\magcir{\raise -2.truept\hbox{\rlap{\hbox{$\sim$}}\raise5.truept
\hbox{$>$}\ }}
\def\rf{\par\noindent\hangindent 20pt}
\catcode`\@=11
\long\def\@makefntext#1{
\protect\noindent \hbox to 3.2pt {\hskip-.9pt
$^{{\ninerm\@thefnmark}}$\hfil}#1\hfill}                %CAN BE USED

\def\@makefnmark{\hbox to 0pt{$^{\@thefnmark}$\hss}}  %ORIGINAL

\def\ps@myheadings{\let\@mkboth\@gobbletwo
\def\@oddhead{\hbox{}
\rightmark\hfil\ninerm\thepage}
\def\@oddfoot{}\def\@evenhead{\ninerm\thepage\hfil
\leftmark\hbox{}}\def\@evenfoot{}
\def\sectionmark##1{}\def\subsectionmark##1{}}

\def\hm{\ \rm {\it h}^{-1} Mpc}
\begin{document}

\centerline{\normalsize\bf Non-Gaussian Dark Matter perturbations}
\centerline{\normalsize\bf and their imprint on the microwave background}

\vspace*{0.6cm}
\centerline{\footnotesize Carlo Baccigalupi}
\baselineskip=13pt
\centerline{\footnotesize\it INFN and Dipartimento di Fisica, 
Universit\`a di Ferrara, Via del Paradiso 12, 44100 Ferrara, Italy;}
\centerline{\footnotesize\it Osservatorio Astronomico di Roma,
Viale del Parco Mellini 84, 00136 Roma, Italy}
\baselineskip=12pt
%\centerline{\footnotesize\it }
\centerline{\footnotesize bacci@oarhp2.rm.astro.it}
\vspace*{0.3cm}
%\centerline{\footnotesize and}
%\vspace*{0.3cm}
%\centerline{\footnotesize MASSIMO PERSIC}
%\baselineskip=13pt
%\centerline{\footnotesize\it Osservatorio Astronomico}
%\centerline{\footnotesize E-mail persic@sissa.it}

%\vfill
\vspace*{0.9cm}
\begin{abstract}
The phases of the Fourier transform of any linear cosmological
perturbation may be random (the Gaussian case)
or not (the non-Gaussian case).
If a non-Gaussian inhomogeneity was
generated during the inflationary era
by some process of very high energy physics, its evolution
may be computed using the usual linear
theory. I focus on the perturbations induced
on the microwave background (CMB) by an 
underlying non-Gaussian
distribution of dark matter (DM). 
Giving the specific example of
an inflationary bubble, I show how non-Gaussian structures 
lying on the last scattering surface (LSS) may give rise 
to ordered patterns in
the CMB anisotropy field.
The next high resolution observations of the CMB sky
may eventually detect these signals.
\end{abstract}

%\vspace*{0.6cm}
\normalsize\baselineskip=15pt
\section{Introduction}

Very high energy physics was dominant
in the very hot phase of the universe.
At $kT\simeq 10^{15}$ GeV the non-zero vacuum
enrgy density of some fundamental field(s) may have
driven a stage of accelerated expansion commonly
known as inflation. This phase allows for the generation
of perturbations in the cosmic density field, seeds of the
structures we observe today, both in the DM distribution
and in the cosmic radiation (see [7] for an extensive
overview). Therefore, the accurate search
for the observable traces of very high energy processes
occurred in the early universe
is very important to gain insight into domains
unaccessible with accelerator physics.

In this contribute I focus on the CMB perturbations
induced by such processes.
The inflationary stage leaves traces in the form of
Gaussian and/or non-Gaussian structures,
described in general in Section II.
In Section III I breafly recall how they are encoded
in temperature perturbations and
represent the seeds for the CMB anisotropies.
Then, in Section IV, I point out that non-Gaussian
DM inhomogeneities on the LSS may give rise to ordered
patterns in the CMB anisotropy field; the usual CMB power
spectrum, carrying the informations about the
cosmological parameters, is not an appropriate tool for
the detection of these signals. I give the specific example
of the CMB anisotropy field caused by a DM
inflationary bubble sitting on the LSS, showing
how it could appear in the forthcoming Planck high resolution
CMB maps. 

	\section{Gaussianity and non-Gaussianity}

At the end of the inflationary stage, the field(s)
involved in the process decay in DM, baryons and radiation,
leaving the cosmic density field $\rho ({\bf x},t)$
perturbed around its mean value $\rho$ by
the traces of the perturbations generated during
inflation: such traces are described through the density
contrast $\delta\rho /\rho =\rho ({\bf x},t)/\rho -1$,
well less than one if the perturbations are linear as
assumed in all the cosmological theories.
In order to distinguish between Gaussianity and non-Gaussianity,
consider the Fourier transform of $\delta\rho /\rho$
written it the following way:
\begin{equation}
\label{gng}
\left({\delta\rho\over\rho}\right)_{\bf k}=
\left|{\delta\rho\over\rho}\right|_{\bf k}
\cdot e^{i\phi_{\bf k}}\ \ ;
\end{equation}
depending on the properties of the phases $\phi_{\bf k}$, you
can have the following distinction:
\begin{itemize}

\item {\bf Gaussian perturbations:}\\
the phases $\phi_{\bf k}$ are random;
$<\left|{\delta\rho\over\rho}\right|_{\bf k}^{2}>$ describes completly
the statistics ($\propto k$ in the simplest inflationary
model). They are generated by the quantum fluctuations
of the field driving the inflationary process during the slow
rolling regime (see [7]).
\item {\bf non-Gaussian perturbations:}
the phases $\phi_{\bf k}$ are not random, but reflect the shape
of the fluctuation; the amplitude 
$\left|{\delta\rho\over\rho}\right|_{\bf k}$
is generally strongly scale dependent (in most cases it ivolves
only a finite set of wavenumbers ${\bf k}$). They are generated
during inflationary phase transitions (defects [5], bubbles
[6]).

\end{itemize}

\section{Temperature perturbations}

At the end of inflation the density
field is perturbed in a generic way described in the previous
section. This induces a corresponding perturbation in the
temeperature field $\delta T/T=T({\bf x},t)/T-1$.
Taking its Fourier transform, each
Fourier mode evolve independently
according to the linear theory; at decoupling ($_{D}$), 
the temperature inhomogeneities become anisotropies 
$(\delta T/ T)(\theta ,\phi )$ in the CMB sky.
Summarizing, the process involves the following steps:
\begin{equation}
{\delta\rho\over\rho}({\bf k},0)
\rightarrow{\delta T\over T}({\bf k},0)\rightarrow
{\rm evolution}\rightarrow{\delta T\over T}({\bf k},t_{D})
\rightarrow
{\delta T\over T}(\theta ,\phi )\ \ .
\label{m}
\end{equation}

The temperature perturbations may be intuitively
imagined as the addition of two components: a photon scattered
at some point characterized by a temperature perturbation
$(\delta T/T)_{0}$ carries the latter plus a Doppler
component simply due to the velocity of the baryon from
which it was scattered, $(\delta T/T)_{D}$.
These two quantities are governed
by the following equations
\begin{equation}\label{hs1}
\ddot{\left({\delta T\over T}\right)_{0}}+
3\dot{\cal R}c_{s}^{2}
\dot{\left({\delta T\over T}\right)_{0}}+
k^{2}c_{s}^{2}{\left({\delta T\over T}\right)_{0}}=
-\ddot{\Phi}-3\dot{\cal R}c_{s}^{2}\dot{\Phi}-
{k^{2}\over 3} \Psi\ \ ,
\end{equation}
\begin{equation}\label{hs2}
\left({\delta T\over T}\right)_{D}=-{3\over k}
\left[\dot{\left({\delta T\over T}\right)_{o}}+\dot{\Phi}\right]
\ ,\ \left({\delta T\over T}\right)_{0}(0)={-\Psi (0)\over 2}\ ,
\ \left({\delta T\over T}\right)_{D}(0)=0\ ,
\end{equation}
and we refer to [4] for a full treatment of them;
here we only remark that the first one looks like a wave equation:
$c_{s}$ is the sound velocity of the system formed by baryons
and photons, coupled by Thomson scattering. Also note that the
system is driven by the gravitational potentials, that in Fourier
transform are proportional to the density contrast:
\begin{equation}
\label{s}
\Phi ,\Psi\propto {\delta\rho\over\rho}\ .
\end{equation}
Finally, due to the same Thomson scattering,
inhomogeneities are exponentially damped below the
Silk damping scale. This is taken into
account by multiplying each component of the perturbation 
by the factor $\exp [-k^{2}/k_{D}^{2}(t)]$, where the
damping scale $k_{D}^{-1}$ is of the order of ten 
comoving Mpc at decoupling.

\vskip .6in

\hskip .3in
\psfig{figure=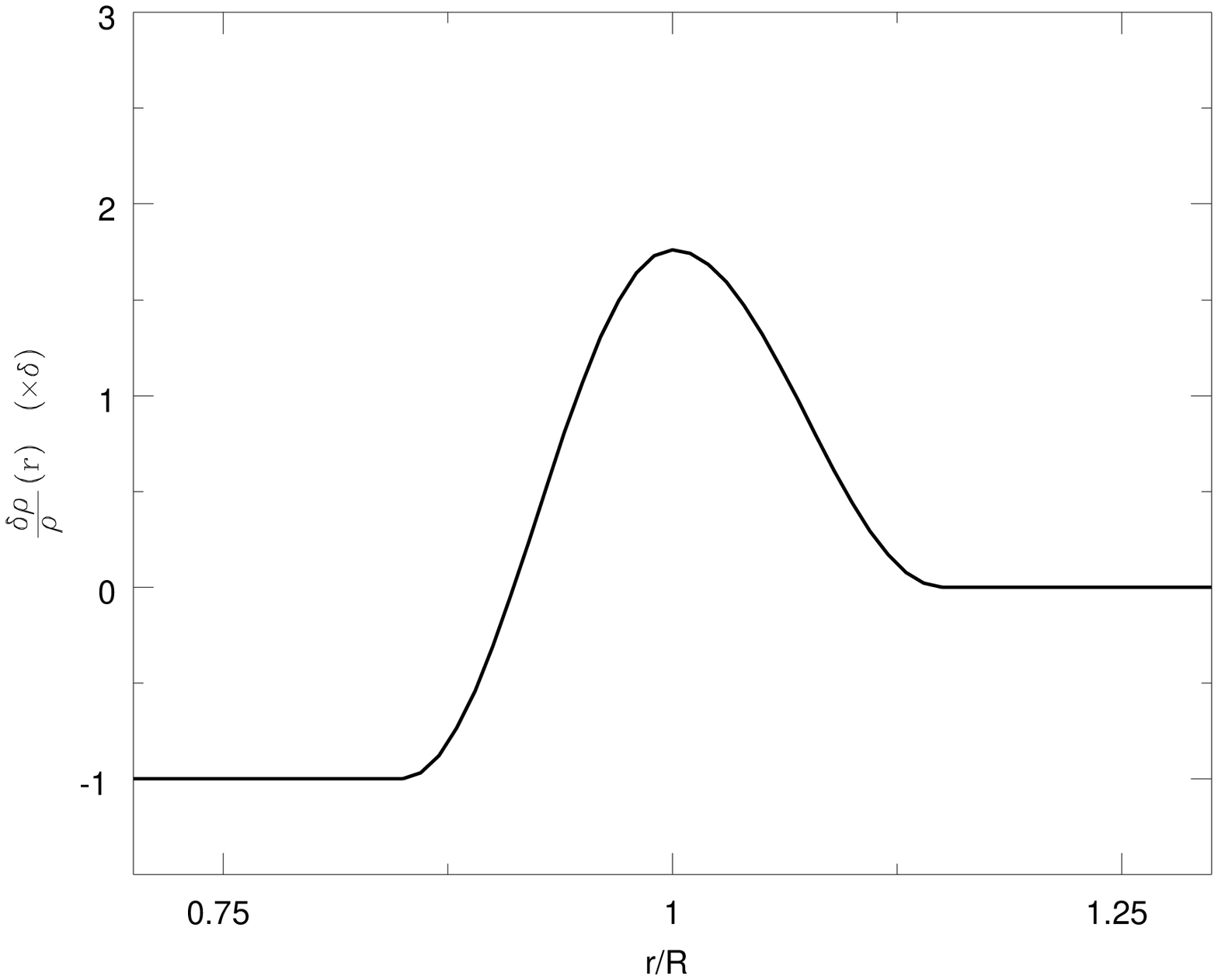,height=2in,width=2in}
\hskip .5in
\psfig{figure=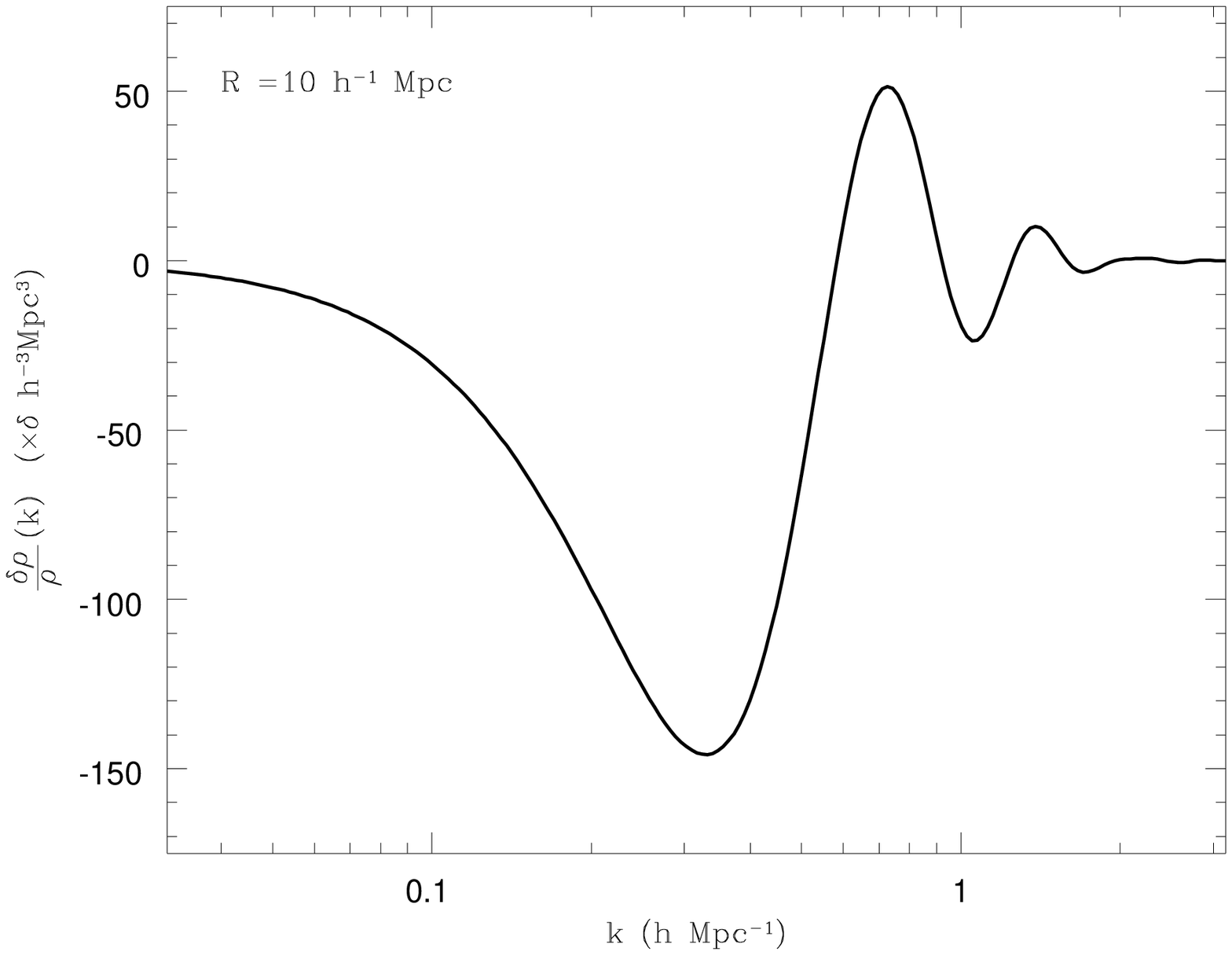,height=2in,width=2in}

%\hskip -3in
%\fcaption{azz}
%\fcaption{azz}

\vskip -1in
\centerline{\footnotesize {\bf Fig.1 (left)}: bubble's density 
contrast radial profile. {\bf Fig.2 (right)}: its Fourier transform.}

\vskip .1in
\section{Perturbation evolution and CMB anisotropies}

As I stated above, the primeval density perturbations
may be Gaussian or not. In the first case, the CMB anisotropies
resulting from the evolution of the system (\ref{hs1},\ref{hs2})
are also Gaussian, and the CMB power spectrum completly describes
the statistics. Instead, this is not true in the second case,
that is the subject here.

A generic non-Gaussian perturbation is in general an ordered
and spatially limited structure.
Consider, just for example, a bubble in the density
field at the end of inflation. Its density contrast may be
easily analitically parameterized [1], and in figure 1
reports the radial density profile: the central cavity and
the compensating shell are well visible; also note the
linear scaling with the central density contrast
$\delta=(\delta\rho /\rho)_{r=0}$.
Due to the spherical symmetry, the Fourier transform only
depends on $k=|{\bf k}|$ and it is real if the frame center
coincides with the center of the bubble, $\phi_{\bf k}=0$ or
$\pi$. Figure 2 reports the amplitude dependence on $k$.
The comoving radius of the bubble is chosen to be $10\hm$ since
it corresponds to the voids and structures observed today
in the modern galaxy catalogues [3].

Each mode $(\delta\rho /\rho )_{k}$ is put as input 
in (\ref{s}) of the
gravitational potentials of the CMB equations at the end
of inflation; the temperature perturbations evolve accordingly
to the initial perturbation and to its own physics. 
At the end
of the evolution all the $k-$modes are backtransformed
to get the temperature perturbation in the real space:
\begin{equation}
\label{f}
\left({\delta T\over T}\right)_{{\bf x},t}=
\int{d{\bf k}\over (2\pi )^{3}}
\left({\delta T\over T}\right)_{{\bf k},t}
e^{i{\bf k}\cdot{\bf x}}\ .
\end{equation}

The general phenomenology for an initial localized inhomogeneity
is the following: the corresponding temperature pertutbation
propagates beyond the initial size, generating waves
travelling outwards with the sound velocity $c_{s}$;
the signature of the underlying DM seed reaches the sound
horizon at the time we are examinating it. This is immediately
evident by looking at figure 3. It represents the radial
profile of $(\delta T/T)_{0}$ as it is at different times
(redshifts), for a bubble with radius $R=20\hm$.
In panel $a$ you can see the initial condition, that remain
unchanged until the horizon reenter occurs (panel $b$); at this
time a first central oscillation occurs, together with
the origin of a positive crest. Soon after (panel $c$)
a second opposite oscillation and the origin of a negative
crest of the sound wave occur. Finally, the decoupling
is reached with a central negative perturbation and
a well visible outgoing sound wave (panel $d$).
\vskip .75in

\hskip 1in 
\psfig{figure=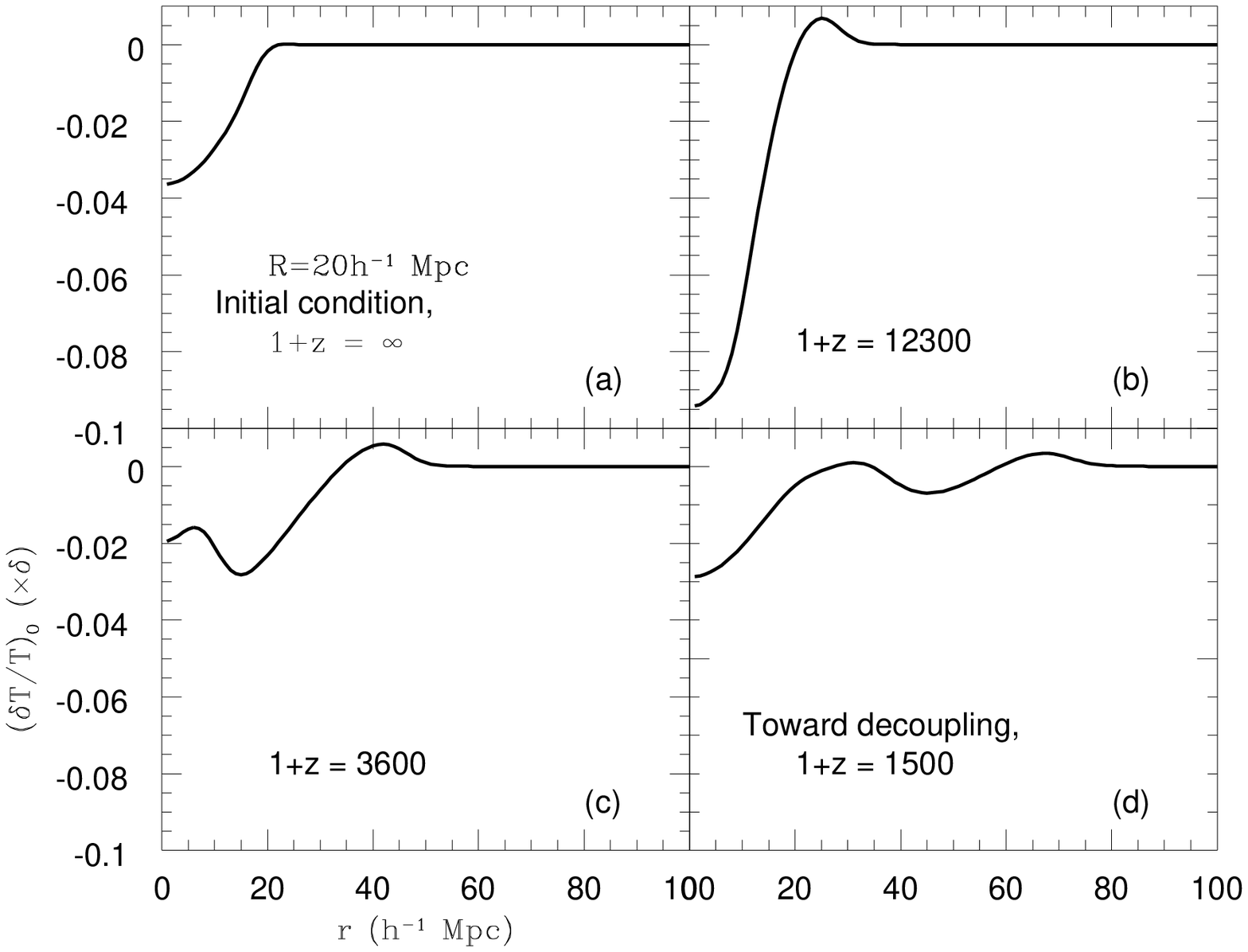,height=2.3in,width=2.5in}

\vskip -1.3in
\centerline{\footnotesize {\bf Fig.3}: Time evolution
of the temperature perturbation induced by a bubble.} 
At decoupling the above perturbations imprint anisotropies
in the CMB sky in the following way.
On a particular direction ${\bf n}$, the CMB anisotropy is
given by
\begin{equation}
\left({\delta T\over T}\right)(\eta)_{\bf n}=\int_{0}^{t_{o}}
\left({\delta T\over T}\right)_{t,{\bf n}}P(t)dt\ \ ,
\label{dtt1}
\end{equation}
where $P(t)$ is the probability for a photon to be last scattered
between times $t$ and $t+dt$. The perturbations examined here are
spatially localized: they have a center and are null beyond
a sound horizon distance from it. Consequentely, in performing
the computation (\ref{dtt1}), we must specify the position
of the perturbation with respect to the LSS.
In the present case it is simply the distance $d$ between
the bubble's center and the LSS. Then the argument of the
integral in (\ref{dtt1}) is
\begin{equation}\label{dtt2}
\left({\delta T\over T}\right)_{({\bf n},d,t)}=\left[
\left({\delta T\over T}\right)_{0}+
\left({\delta T\over T}\right)_{SW}+
\left({\delta T\over T}\right)_{D}\cos{\theta_{\bf n}}
\right]_{({\bf n},d,t)}\ ,
\end{equation}
where the cosine of course accounts for the direction dependence
of the Doppler effect and $(\delta T/T)_{SW}$
is the Sachs-Wolfe contribute given by the gravitational
potential at the last scattering point
(see [4]).

Figure 4 reports the CMB anisotropies from bubbles lying
on the LSS; $\theta$ is the angle between the line of sight
and the bubble's center direction.
In each panel, three different lines show the variation
of the angular dependence of $\delta T/T$ with the relative positions
of the LSS with respect to the bubble's center: $d=0,\pm 20\hm$;
the central negative spots and the outgoing waves are well
evident in both panels.
\vskip .75in

\hskip 1in 
\psfig{figure=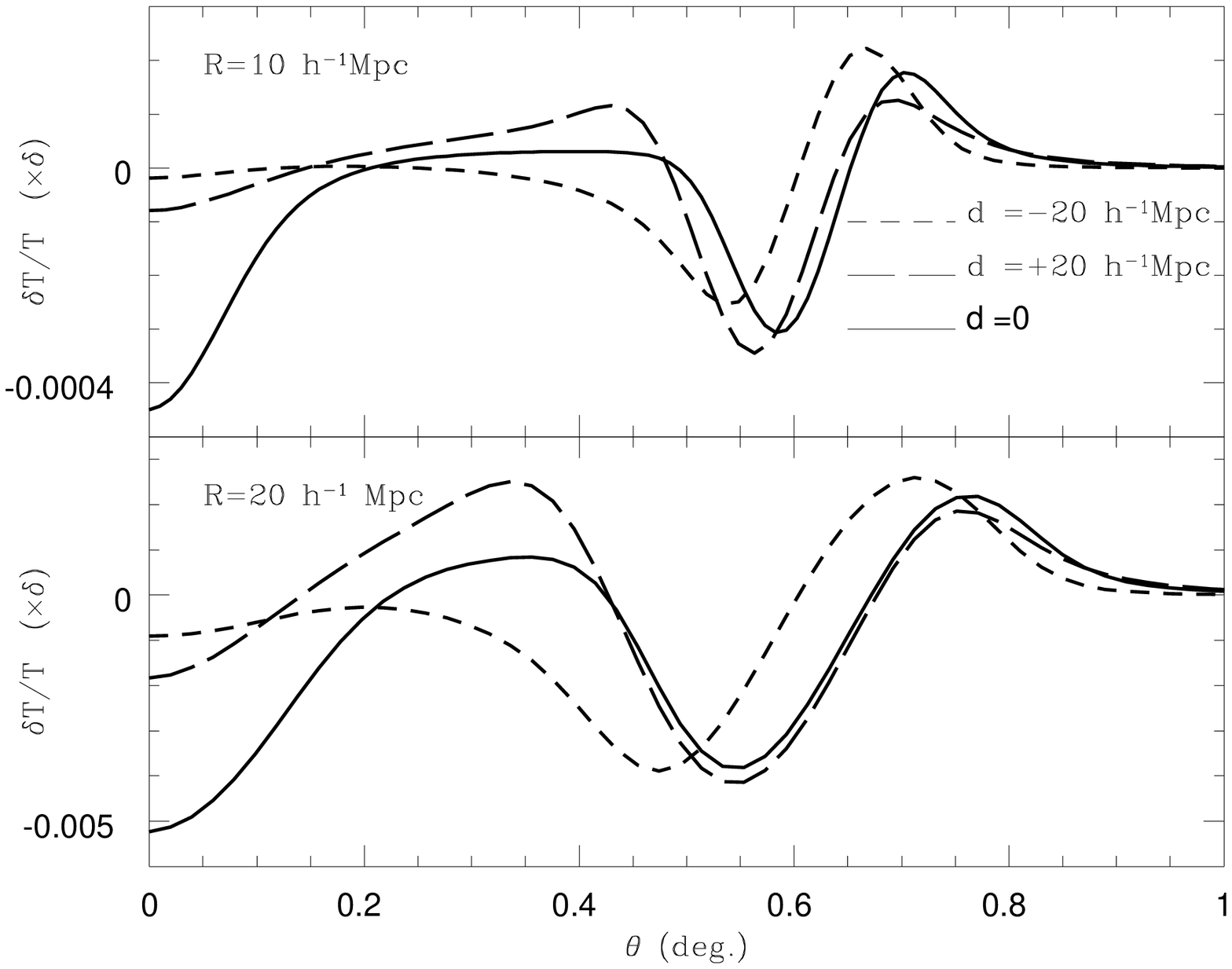,height=2.5in,width=2.5in}

\vskip -1.3in
\centerline{\footnotesize {\bf Fig.4}: CMB anisotropies
induced by a bubble lying on the LSS.}

Finally look at figure 5, bottom panel;
this is a portion of microwave sky in which a bubble with
$R=20\hm$ and $\delta =10^{-2}$ lies exctly on the LSS ($d=0$);
the angular coordinates are in primes. In the general case, the
non-Gaussian structures like the ones analyzed
here are embedded in a Gaussian perturbation field
filling the whole space [2]: then the figure reports
the addition of the Gaussian CMB anisotropy plus the component
given by the bubble. The circular imprint is well evident.
Note that the DM bubble is in the central small dark spot,
while the rings surrounding it are caused by the
outgoing sound waves; consequentely, the signal has
the angular scales of the sound horizon at decoupling,
that is nearly $1^{o}$ in the sky. 
Just like a pebble in a pond, the initial small bubbly
perturbation is propagating beyond the 
initial scale, reaching the scale of the sound horizon at the time 
in which we are examinating it.
In the upper panel,
$\delta$ is half: in this case the Gaussian noise is
quite dominant even if the undelying non-Gaussianity
affects the statistics and may be searched with
more sophisticated tools than the human eye.

If these structures are in a number sufficient to affect
the structure formation, they are expected to perturb
the whole sky CMB power spectrum. In [2] we
have analyzed in detail this case, showing that for a
large number of possible statistics a distinctive peak
at angular scales corresponding to $10^{'}$ arises.
This is not true for all the cases simply because when
the statistics is not Gaussian the power spectrum
does not fix univoquely the perturbation field.
Thus, the ultimate imprint of the structures analyzed
here has to be searched in higher order statistics. 

Sub-degree full sky CMB maps shall be performed in the near
future by the MAP and Planck experiments; they are
consequentely the most appropriate observations for the
eventual detection of the signals treated here.

%\vskip .75in

\hskip 1in 
\psfig{figure=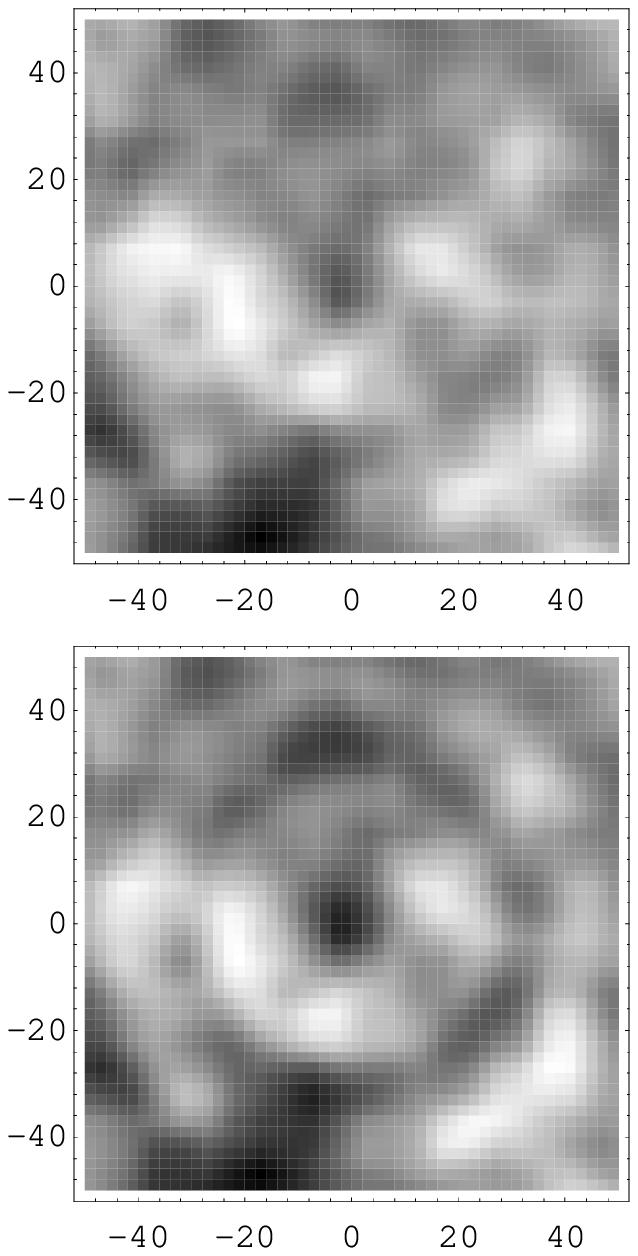,height=5.3in,width=4.3in}

\vskip -2.3in
\centerline{\footnotesize {\bf Fig.5}: a portion of
microwave sky generated by a bubble with $\delta =10^{-2}$
(bottom)}
\centerline{\footnotesize and $\delta =5\cdot 10^{-3}$ (top) 
lying on the LSS plus the standard Gaussian perturbations.}

\section{References}
\bigskip

\rf{[1] Baccigalupi C. 1998 {\it Ap.J.} {\bf 496} in press.}
\rf{[2] Amendola L., Baccigalupi C. \& Occhionero F. 1998 
{\it Ap.J.Lett.} {\bf 492} in press.}
\rf{[3] El-Ad H., Piran T. \& da Costa L. N., 1997 {\it MNRAS}
{\bf 287} 790.}
\rf{[4] Hu W. \& Sugiyama N. 1995 
{\it Ap.J.} {\bf 444}, 489}
\rf{[5] Pen U., Seljak U. \& Turok N. 1997 {\it Phys.Rev.Lett.} {\bf 79}, 
1611}
\rf{[6] Occhionero F. \& Amendola L.  1994 
{\it Phys.Rev.D} {\bf 50} 4846.}
\rf{[7] Padmanabhan T., Structure 
formation in the  universe, {\it Cambridge university press} 1993.}

\end{document}